\documentclass{PoS}

\usepackage{bm}
\usepackage{mathptmx}
\usepackage{amsmath}
\usepackage{bm}
\usepackage{textcomp}

\DeclareMathAlphabet\mathbfcal{OMS}{cmsy}{b}{n}

\title{The Tunka Radio Extension: reconstruction of energy and shower maximum of the first year data}

\ShortTitle{Tunka-Rex: reconstruction of energy and shower maximum}

\author{
\speaker{D.~Kostunin}$^{1}$, P.A.~Bezyazeekov$^{2}$, N.M.~Budnev$^{2}$, O.A.~Gress$^{2}$, A.~Haungs$^{1}$, R.~Hiller$^{1}$, T.~Huege$^{1}$, Y.~Kazarina$^{2}$, M.~Kleifges$^{3}$, E.N.~Konstantinov$^{2}$, E.E.~Korosteleva$^{4}$, O.~Kr\"omer$^{3}$, L.A.~Kuzmichev$^{4}$, N.~Lubsandorzhiev$^{4}$, R.R.~Mirgazov$^{2}$, R.~Monkhoev$^{2}$, A.~Pakhorukov$^{2}$, L.~Pankov$^{2}$, V.V.~Prosin$^{4}$, G.I.~Rubtsov$^{5}$, F.G.~Schr\"oder$^{1}$, R.~Wischnewski$^{6}$, A.~Zagorodnikov$^{2}$
- 
Tunka-Rex Collaboration \\
\llap{$^1$} Institut f\"ur Kernphysik, Karlsruhe Institute of Technology (KIT), Germany\\
\llap{$^2$} Institute of Applied Physics ISU, Irkutsk, Russia\\
\llap{$^3$} Institut f\"ur Prozessdatenverarbeitung und Elektronik, Karlsruhe Inst.~of Tech.~(KIT), Germany\\
\llap{$^4$} Skobeltsyn Institute of Nuclear Physics MSU, Moscow, Russia\\
\llap{$^5$} Institute for Nuclear Research of the Russian Academy of Sciences, Moscow, Russia\\
\llap{$^6$} DESY, Zeuthen, Germany\\
E-mail: \email{dmitriy.kostunin@kit.edu}       
}

\abstract{
Since its commissioning in autumn 2012, Tunka-Rex, the radio extension of the air-Cherenkov detector Tunka-133, performed three years of air shower measurements. 
Currently the detector consists of 44 antennas connected to air-Cherenkov and scintillator detectors, respectively, placed in the Tunka valley, Siberia. 
Triggered by these detectors, Tunka-Rex measures the radio signal up to EeV-scale air-showers. 
This configuration provides a unique possibility for cross-calibration between air-Cherenkov, radio and particle techniques. 
We present reconstruction methods for the energy and the shower maximum developed with CoREAS simulations, which allow for a precision competitive with the air-Cherenkov technique. 
We apply these methods to data acquired by Tunka-Rex in the first year which we use for cross-calibration, and we compare the results with the reconstruction of the energy and the shower maximum by Tunka-133, which provides also a reconstruction for the shower core used for the radio reconstruction.
Our methods have shown that the atmospheric depth of the shower maximum ($X_\mathrm{max}$) can be reconstructed with a precision of better than 40~g/cm$^2$ for high quality events, in some cases even when only three antenna stations have signal.
The energy precision is comparable with the air-Cherenkov precision of 15\%.
Soon the results will be checked with the independent data of the second year.
}

\FullConference{The 34th International Cosmic Ray Conference,\\
		30 July- 6 August, 2015\\
		The Hague, The Netherlands}

\begin{document}

\section{Introduction}
Radio detection of high-energy cosmic rays recently has become a competitive technique.
Provided a simple external trigger it reaches a duty cycle around-the-clock.
A first generation of modern radio experiments~\cite{Falcke:2005tc,Ravel:2012zz} has opened the prospects cosmic-ray detection with energies higher than 0.1~EeV.
The dense radio-astronomical observatory LOFAR demonstrated good agreement between the amplitude footprint predicted by CoREAS and real measurements~\cite{Buitink:2014eqa}.
After the confirmation that the radio signal is sensitive to the air-shower parameters energy and shower maximum, the main remaining question is the applicability to large scales the radio technique for the detection of ultra-high energy cosmic rays.
Currently there are two experiments aiming to answer this question: AERA~\cite{icrc2015Schulz} and Tunka-Rex~\cite{hiller_NIMA}.

Tunka-Rex is the radio extension of the air-Cherenkov detector Tunka-133~\cite{Prosin:2014dxa} placed in Siberia, near Lake Baikal.
The host experiment Tunka-133 is successfully taking data since 2009.
It operates during winter in moonless nights, reaching a total duty cycle of less than 500~h/year.
The detector array consists of 19 clusters placed on 1~km$^2$ area, each equipped with 7 photomultiplier tubes (PMT) with spacing 80 m.
Later it was extended by 6 externals clusters increasing the effective detection area up to 3~km$^2$.
Recently an additional array of scintillators, Tunka-Grande, and air-Cherenkov extensions under the name of TAIGA have been installed~\cite{Budnev:2014zna}.

The radio extension (Tunka-Rex) consists of 44 antennas, 25 of them are connected to Tunka-133 cluster centers.
Another 19 are connected to the scintillator extension currently under commissioning.
The Tunka-Rex detector works in slave mode, i.e. it receives trigger from the air-Cherenkov array; all events are recorded in parallel.
This provides the ultimate possibility of cross-calibrating the radio and the air-Cherenkov techniques.
Since Tunka-133 has a high resolution (15\% for the energy and 28~g/cm$^2$ for the shower maximum reconstruction), Tunka-Rex has the potential to determine the precision of the radio technique.

Tunka-Rex started its data acquisition in October 2012.
Until now, three seasons of data acquisition were completed (2012/2013, 2013/2014 and 2014/2015).
Our collaboration decided to follow a semi-blind analysis paradigm: energy and shower maximum reconstruction by Tunka-133 is blinded starting from the second season.
In the present work we show the results of the first season used for tuning and testing the reconstruction method.
These methods are used to predict the energy and shower maximum for the second season enabling an independent cross-check.

\section{Reconstruction of air-shower parameters}
For the reconstruction of air-shower parameters we use a common method exploiting the lateral distribution of amplitudes.
In principle, if one knows the shower axis and core position, this method requires only the amplitude at two antenna stations for the energy and shower maximum reconstruction, since mainly these two shower parameters are defining the lateral distribution.
The amplitude is sensitive to the energy, and the shape of the lateral distribution to the shower maximum (see, for example, Ref.~\cite{Huege:2008tn}).
For the radio detection, the main difficulty rises from two features of the generation of radio emission. 
First, the interference of the geomagnetic effect~\cite{Kahn:1966} and the charge excess~\cite{Askaryan:1962} introduces an azimuthal asymmetry in the lateral distribution: the distribution becomes two-dimensional.
The second effect comes from the features of the signal propagation in a medium (atmosphere) with refractive index $n_r > 1$: the lateral distribution contains Cherenkov-like features.
Taking these features into account, we developed an appropriate methods to describe measured lateral distributions.
Our study is based on CoREAS~\cite{Huege:2013vt} simulations and described in Ref.~\cite{Kostunin:2015taa}.

\subsection{Asymmetry}
In Ref.~\cite{Buitink:2014eqa}, it was shown that the charge excess contribution to the radio emission leads to significant complication of lateral distribution.
Nevertheless, this distribution can be fitted with a two-dimensional function~\cite{Nelles:2014xaa}.
This solution is very general, but hardly applicable to sparse detectors operating near threshold, i.e. a typical event contains only 3-5 antenna stations with signal.
We provide a simpler approximation of the lateral distribution and parameterize it with a one-dimensional function reducing the number of free parameters.

The squared signal amplitude at a certain distance has the form
\begin{equation}
\mathcal{E}^2 = \mathcal{E}_0^2\left( (\sin\alpha_{\mathrm g} + \varepsilon\cos\phi_{\mathrm g})^2 + \varepsilon^2\sin^2\phi_{\mathrm g} \right),
\label{squared_amplitude}
\end{equation}
where $\varepsilon$ is the asymmetry defined as fraction of strengths of the Askaryan relative to the geomagnetic, $\alpha_{\mathrm g}$ is the geomagnetic angle (angle between vector of magnetic field and shower axis), $\phi_{\mathrm g}$ is the azimuth of an antenna station in the shower plane.

As we can see, the one-dimensional LDF transforms itself to a two-dimensional one when taking into account the contribution from the charge excess phenomena: $\varepsilon > 0$.
To reduce the number of dimensions back to one, we define a special operator $\mathsf{\hat K}$ eliminating the azimuthal dependence
\begin{equation}
\mathsf{\hat K} = \sqrt{\varepsilon^2 + 2\varepsilon\cos\phi_{\mathrm g}\sin\alpha_{\mathrm g} + \sin^2\alpha_{\mathrm g}}\,,
\end{equation}
The remaining question is the value of the asymmetry $\varepsilon$, which should be used for the correction.
In Ref.~\cite{Kostunin:2015taa} it was shown with CoREAS simulations that a constant value $\varepsilon = 0.085$ is sufficient for the correction of the lateral distribution function for the situation of Tunka-Rex.

\subsection{Lateral distribution function (LDF)}
As mentioned, due to the refractive index a Cherenkov ring is present in the lateral distribution~\cite{deVries:2013dia,Nelles:2014dja}.
Thus, we use the simplest function which can describe the Cherenkov ring~\cite{kalm}
\begin{equation}
\mathcal{E}(r) = \mathcal{E}_{r_0} \exp(a_1(r-r_0) + a_2(r-r_0)^2)\,,
\label{general_1dim_ldf}
\end{equation}
where $r$ is the distance to shower axis.
A problem rising from this parameterization is connected to the strong correlation between the parameters $a_1$ and $a_2$.
When the number of fitted points is close to the number of fit parameters, in addition to large error bars, the fit can converge to false minima, or give large uncertainties for parameters with strong correlation.
Unfortunately, most of the Tunka-Rex events satisfy these conditions.
For this reason we decided to reduce the number of free parameters in the LDF by fixing the parameter $a_2$ to a value depending on zenith and energy
\begin{eqnarray}
&&a_2(\theta, E_\mathrm{pr}) = a_{20}(E_\mathrm{pr}) + a_{21}(E_\mathrm{pr})\cos\theta\,,\\
&&a_{20} = a_{200} + a_{201}E_\mathrm{pr}\,,\,\, a_{21} = a_{210} + a_{211}E_\mathrm{pr}\,,
\label{a2param}
\end{eqnarray}
with $a_{200} = 0.19\cdot 10^{-5}$~m$^{-2}$, $a_{201} = -1.63\cdot 10^{-5}$~m$^{-2}/$EeV, $a_{210} = -3.45\cdot 10^{-5}$~m$^{-2}$, $a_{211} = 2.44\cdot 10^{-5}$~m$^{-2}/$EeV obtained by the CoREAS simulations.
The primary energy $E_\mathrm{pr}$ for this parameterization is estimated with a simple exponential LDF after correction for the asymmetry.
The final energy estimator is described in the next section.

\subsection{Energy and shower maximum reconstruction}
The detailed explanation of formulas used for the reconstruction of air-shower parameters is given in Ref.~\cite{Kostunin:2015taa}: here we only give the parameters obtained from the model.
The energy can be reconstructed by probing the signal amplitude at a defined distance $r_{\mathrm e}$
\begin{equation}
\label{energy_rec} E_{\mathrm{pr}} = \kappa_L \mathcal{E}(r_{e})\,,
\end{equation}
where parameters have the following values: $r_\mathrm{e} = 
120\mbox{ m}$, $\kappa_L = 884\mbox{ EeV}/(\mbox{V/m})$\footnote{Here we use $\kappa_L$ which differs from the value $\kappa = 422\mbox{ EeV}/(\mbox{V/m})$ in Ref.~\cite{Kostunin:2015taa}.
This is because we simplified the equation of Ref.~\cite{Kostunin:2015taa} by setting the exponent $b = 1$ instead of $0.93$.
}.

For $X_\mathrm{max}$ reconstruction we use the parameterization
\begin{equation}
X_{\mathrm{max}} = X_{\mathrm{det}} / \cos\theta - (A + B\log(a_1(r_\mathrm{x}) + \bar b))\,,
\end{equation}
where $X_{\mathrm{det}} = 955$~g/cm$^2$ is the atmospheric depth of the detector. The parameters have the following values: $r_\mathrm{x} = 180\mbox{ m}$, $A = -1864 \mbox{ g/cm}^2$, $B = -566 \mbox{ g/cm}^2$, $\bar{b} = 0.005 \mbox{ m}^{-1}$.

\section{Event reconstruction and comparison with Tunka-133}
We perform the cross-calibration for the events of 2012/2013, and make the prediction for the data of the season 2013/2014, which is still blinded.
After unblinding we will compare the prediction given by Tunka-Rex and Tunka-133, and present the conclusion for the precision of the air-shower reconstruction.
These results will be published in another paper soon.
In the present work we present results of the first season in comparison with Tunka-133 reconstruction for the energy and shower maximum.

\subsection{Data acquisition and signal selection}
All events are triggered by air-Cherenkov detector, the radio detector is read out in parallel.
As main software for data analysis we use a modified version of the radio extension of the Auger Offline framework~\cite{Abreu:2011fb}.
For the reconstruction of the radio signal we apply antenna patterns, and hardware responses based on calibration measurements~\cite{icrc2015Hiller}.
As result, we obtain signal traces for each antenna station in the units of the electrical field.
We require at least three stations with signals passing a signal-to-noise (SNR) ratio cut: $\mathrm{SNR} = S^2/N^2 > 10$, where $S$ is a signal amplitude and $N$ is the RMS of noise calculated in a noise window.
With simulations we parameterized the impact of the noise on the measured amplitude and correct for this.

\begin{figure}[h!]
\begin{center}
\includegraphics[width=1.0\linewidth]{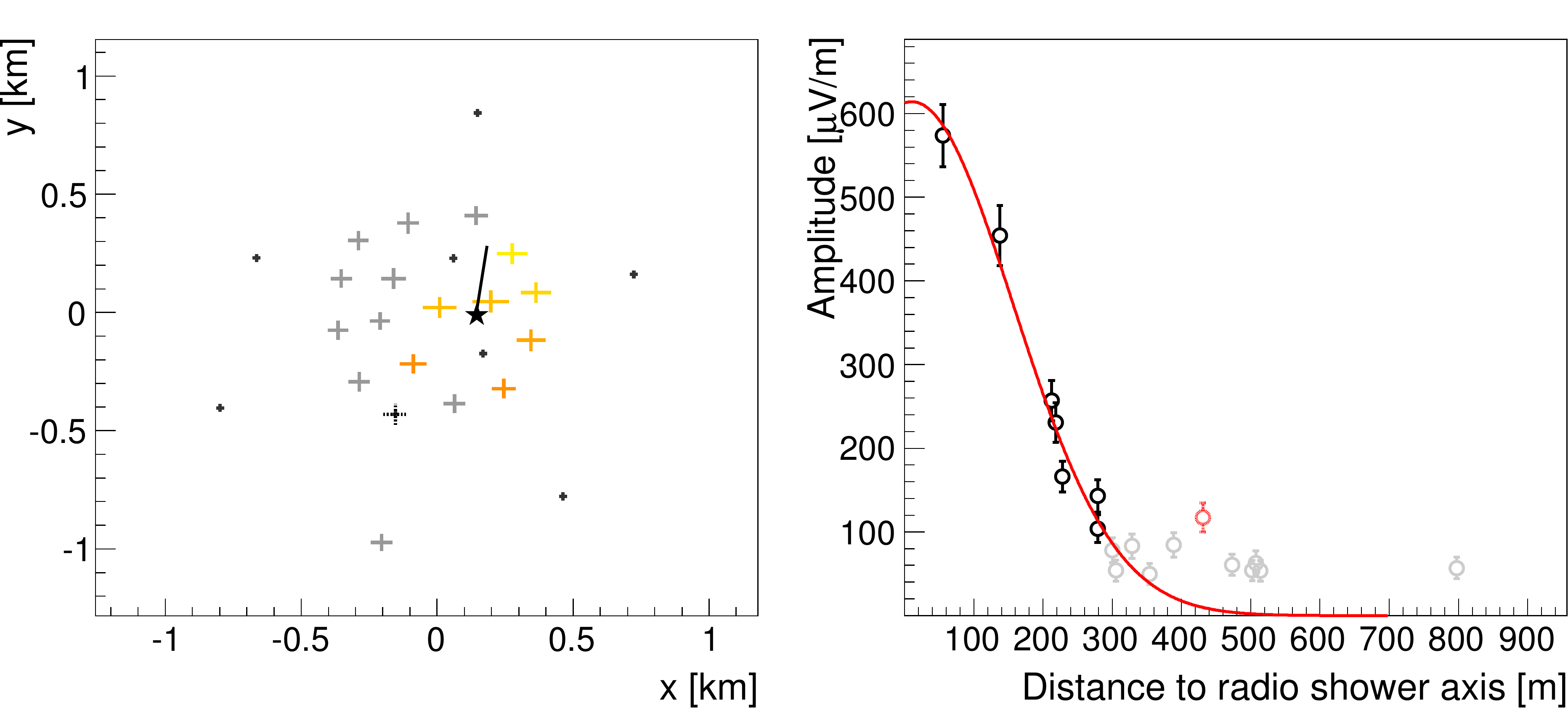}
\end{center}
\caption{
Example event reconstructed by Tunka-Rex.
{\it Left:} Footprint of the air-shower. The color code shows the arrival time. Points show clusters without antenna in 2012/2013.
{\it Right:} Lateral distribution with fit. Grey points show the antennas not passing the SNR cut, the dotted point is a rejected antenna with false signal.
}
\label{eventsample}
\end{figure}

Using the time information of these stations we can reconstruct the air-shower arrival direction and compare it with the reconstruction given by Tunka-133.
Events with big deviation ($> 5^\circ$) in arrival reconstruction are considered as false and rejected.
As last step we remove outliers from the lateral distribution and fit this distribution with our parameterization, if it still has at least three antenna stations with signal.
An example event is shown in Fig.~\ref{eventsample}.

In the season 2012/2013, the Tunka facility had effectively 280~h of measurement time.
The full reconstruction of Tunka-133 is available only for events with zenith angles $\theta \le 50^\circ$ due to design features.
Thus, for the Tunka-Rex cross-calibration we used only these events.
After the cuts described above we have 91 events.

\subsection{Energy reconstruction}
\begin{figure}[t]
\begin{center}
\includegraphics[width=0.95\linewidth]{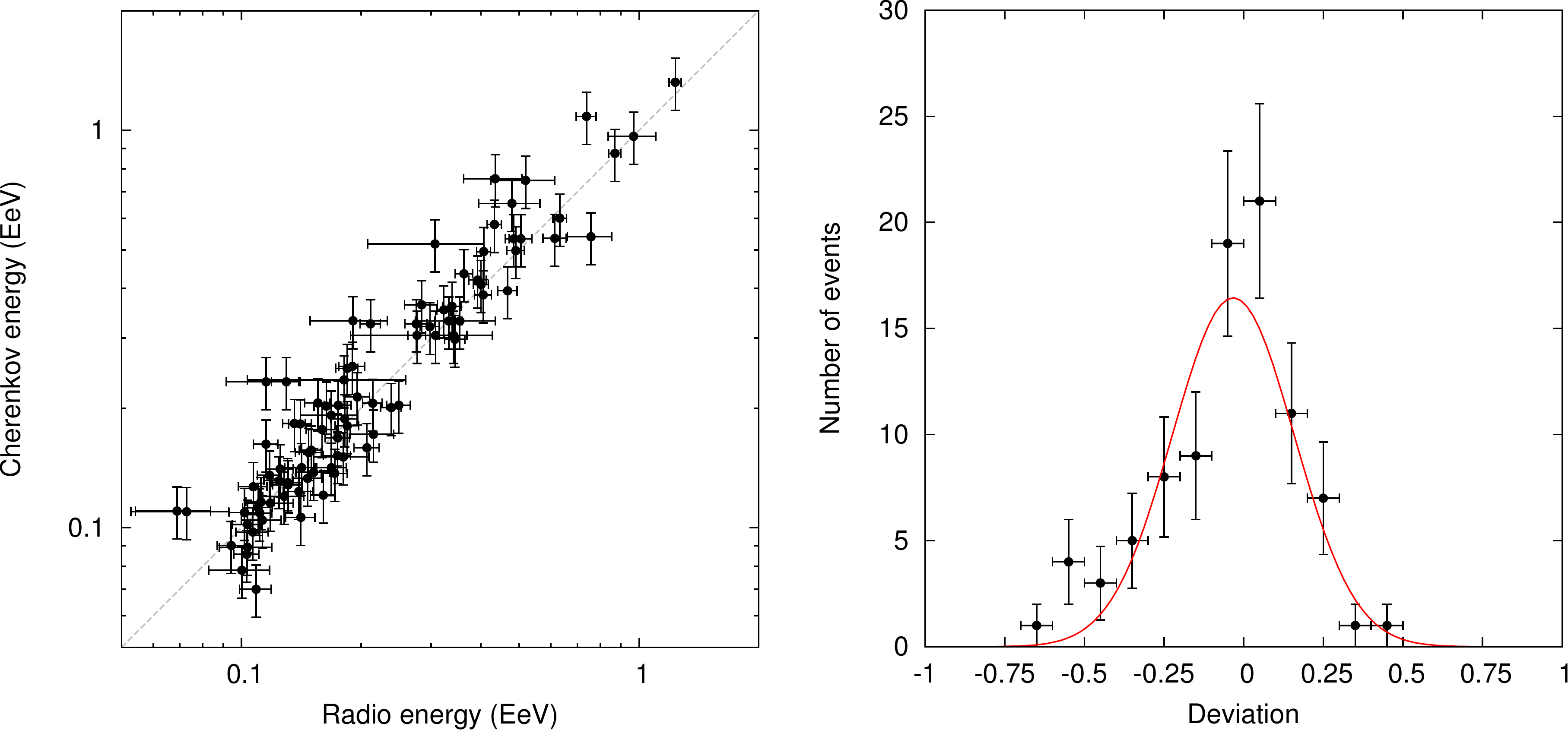}
\end{center}
\caption{
Correlation between the energy reconstructed by Tunka-Rex and by Tunka-133, and histogram of relative deviation.
The deviation is calculated as energy difference divided by average.
}
\label{energy}
\end{figure}

We present the results of the energy reconstruction given by formula~(\ref{energy_rec}).
The spread between values given by Tunka-Rex and Tunka-133 is $19 \pm 3$\% when using all 91 events (see Fig.~\ref{energy}).

\subsection{Shower maximum reconstruction}
\begin{figure}[t]
\begin{center}
\includegraphics[width=0.95\linewidth]{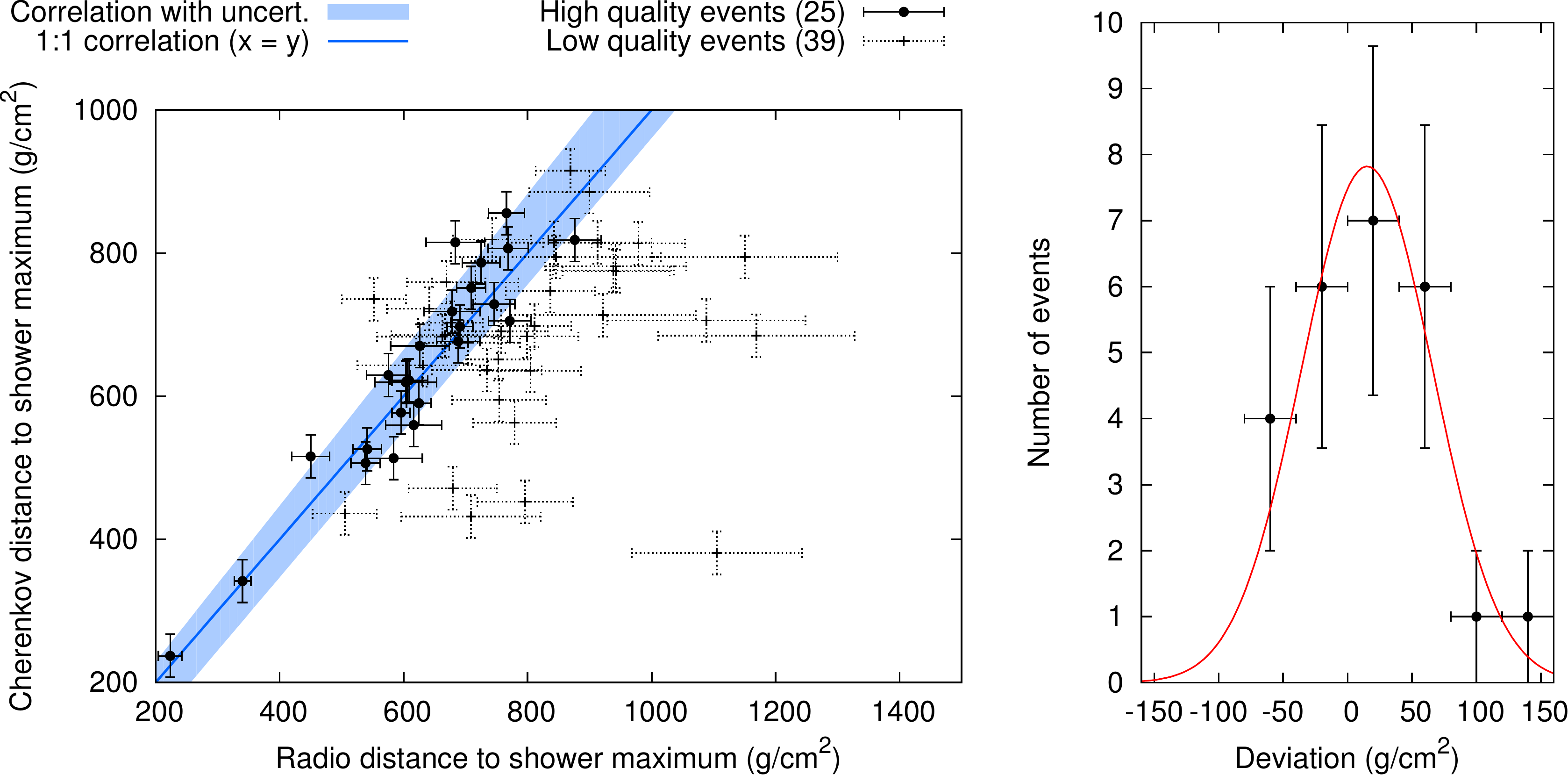}
\end{center}
\caption{
Correlation between the distance to the shower maximum reconstructed by Tunka-Rex and by Tunka-133m and histogram of the relative deviation.
Both reconstructions are in agreement within the uncertainties.
The mean deviation is $51 \pm 7$ for the high and $130 \pm 12$~g/cm$^2$ for the low quality cuts.
The deviation is calculated as $X_\mathrm{max}$ difference.
}
\label{xmax}
\end{figure}

The reconstruction of the shower maximum is connected to the slope of the lateral distribution.
Dense detectors have shown good sensitivity to the shower maximum~\cite{Buitink:2014eqa,Nelles:2014xaa}, but the reconstruction with sparse detectors as Tunka-Rex is more complicated.
Events with small number of antenna stations can be used to reconstruct the shower maximum, only if the fitted slope (parameter $a_1$ of Eq.~(\ref{general_1dim_ldf})) has a small uncertainty.
However, a small number of antenna stations means that the lateral distribution has points close to each other, which leads to an underestimation of the slope, i.e. an overestimation of the distance to the shower maximum (see dotted events in Fig.~\ref{xmax}).
Since we probe the slope at the distance of 180~m, we exclude all events which have not antennas at a distance $\ge 200$~m to the shower axis, thus
cut removing 27 events.
The fitting uncertainty of $a_1$ propagates to the uncertainty of the shower maximum reconstruction $\sigma_{X_\mathrm{max}}$.
We define a high quality cut for this uncertainty $\sigma_{X_\mathrm{max}} < \sigma_{X_\mathrm{max}}^{\mathrm{th}}$.
For the present analysis we set $\sigma_{X_\mathrm{max}}^{\mathrm{th}} = 50$~g/cm$^2$.
After this cut 25 events survive (bold points in Fig.~\ref{xmax}).
The mean deviation between Tunka-133 and Tunka-Rex for this quality data set is $51 \pm 7$ g/cm$^2$.
Decreasing the cut threshold $\sigma_{X_\mathrm{max}}^{\mathrm{th}}$ we further increase the precision of Tunka-Rex losing more events.
The best result reached with our statistics is $36\pm 7$~g/cm$^2$ for 15 events.
This means that the uncertainty $\sigma_{X_\mathrm{max}}$ is consistently estimated.
If the Tunka-Rex uncertainty for $X_\mathrm{max}$ would be equal to the Tunka-133 uncertainty we would expect a mean deviation of $\sqrt{2} \cdot 28 \approx 40$~g/cm$^2$. 

\section{Conclusion}
Tunka-Rex is an excellent experiment to provide a benchmark for the radio detection of cosmic rays.
The detector is a sparse array of antennas, and operates near the detection threshold.
Most events have low SNR and contain only about 3-5 antennas with signal.
On the other side, the experiment is hosted by a precise air-Cherenkov array, which reduces the systematic uncertainty for cross-calibration.

Our study has shown that taking into account all significant effects of the radio emission allows to work under these strong restrictions.
One of the most important questions in this topic is the precision for the reconstruction of the shower maximum.
We have shown that a sparse array with spacing of 200~m is able to reconstruct the shower maximum with uncertainty better than 40~g/cm$^2$ even for some events containing only three antennas, if the event contains antennas with signal further than 200~m from the shower axis for a high-quality selection.
Tunka-Rex within its energy range and detector size has uncertainty of about 50~g/cm$^2$ after cross-calibration with Tunka-133, which has a precision of 28~g/cm$^2$.
Taking all the aforesaid into consideration, one can state, that radio detection is economically feasible for ultra-high energy cosmic rays detection.

Tunka-Rex recently has been extended with additional 19 antennas connected to Tunka-Grande, the scintillator extension of the Tunka facility.
Triggering by the particle detectors will increase the duty cycle by an order of magnitude.
This configuration provides a possibility for cross-calibration between three air-shower detection techniques, as well as an increase of the statistics of the Tunka facility in the energy range around 1 EeV.
\section*{Acknowledgement}
Tunka-Rex has been funded by the German Helmholtz association (grant HRJRG-303) and
supported by the Helmholtz Alliance for Astroparticle Physics (HAP), as well as KCETA.
This work was supported by the Russian Federation Ministry of Education and Science
(agreement 14.B25.31.0010, zadanie 3.889.2014/K) and
the Russian Foundation for Basic Research
(Grants 12-02-91323, 13-02-00214, 13-02-12095, 14002-10002).

\end{document}